# Unified Flow and Thermal Law-of-the-wall Complete Formulations for Type-A Turbulent Boundary Layers

Duo Wang[1], Heng Li[1], Ting Yu[1], Bochao Cao[1] and Hongyi Xu[1]

[1] *Department of Aeronautics and Astronautics,*
*Fudan University, Shanghai, China, 200433*
*E-mail address: Hongyi_Xu@fudan.edu.cn*

In-depth analyses of existing direct numerical simulations (DNS) data from various sources supported a logical and important classification of generic turbulent boundary layers (TBL), namely Type-A, -B and -C TBL, based on distribution patterns of time-averaged wall-shear stress. Among these types, Type-A TBL and its related law, as represented by the DNS data of turbulence on a zero-pressure-gradient semi-infinite flat-plate, was investigated in terms of analytical formulations of velocity independent on Reynolds ($Re$) number. With reference to the analysis from von Karman in developing the conventional law-of-the-wall, the current study first physically distinguished the time-averaged local scale used by von Karman from the ensemble-averaged scale defined in the paper, and then derived the governing equations with the $Re$-independency under the ensemble-averaged scales. Based on indicator function (IDF) and TBL thickness, the sublayer partitions were rigorously defined. The analytical formulations for entire TBL, namely the complete law-of-the-wall, were established, including the formula in inner, buffer, semi-logarithmic (semi-log) and wake layer. The researches were featured by introducing the general damping and enhancing functions (GDF and GEF) and applying these functions to both linear and logarithmic coordinates. These law formulations were proved uniform and consistent in time-averaged local and ensemble-averaged scales, which were validated by the existing DNS and experiment data. Based on the similarity of relevant properly-scaled governing equations, the law formulations were logically reasoned being applicable to the temperature in Type-A thermal TBL. The findings advance the current understandings of the conventional TBL theory and its well-known foundations of law-of-the-wall.

Over the past century, searching for the statistical law in wall-bounded turbulence has always been a persistent effort for fluid mechanics community. The law-of-the-wall [1] is a milestone in Prandtl's turbulent boundary layer (TBL) theory [2]. The law was found through an in-depth study of the measurements for flat-wall turbulence and a unique analysis using the time-averaged local frictional velocity scale [1]. So far, the law has been validated by experiments [3], theoretical methods [4], and modern Direct Numerical Simulations (DNS) for simple geometry TBLs [5-12], such as the TBLs in flat plate, channel or circular pipe. As DNS studies extended to more wall-bounded turbulences, the growing DNS data permitted to investigate TBL's using different scales. Within the context, Cao & Xu [13] first distinguished the time-averaged local frictional velocity $u_\tau$ from the time-space-averaged, or ensemble-averaged frictional velocity $\overline{u_\tau}$, and pointed out the necessity to classify generic TBLs into three types, namely Type-A, -B and -C TBL, according to the distribution patterns of $u_\tau$. Consequently, the inner-layer law formulation was derived for Type-B TBL and was validated by the DNS-guided near-wall integration of governing equation. Inspired by von Karman and Xu's work, the current work used the $\overline{u_\tau}$ to rescale the Type-A TBL governing equations. The rescaled equations were proved Reynolds number ($Re$) independent, which formed a theoretical base for exploring the universally applicable law for Type-A TBL. With the guidance of rescaled DNS data, a physics-oriented analytical design was conducted by making use of the general damping and enhancing functions (GEF and GDF) [13] to correct the tradition linear law ($u^+ = y^+$) and semi-log law ($u^+ = 1/\kappa \ln y^+ + C$) under both linear and logarithmic coordinates. The complete law-of-the-wall for Type-A TBL was derived, including the analytical formula in the inner, buffer, semi-log and wake sublayer under time-averaged local and ensemble-averaged scales, respectively. The law was rigorously validated by the DNS data from Schlatter, Orlu [8][9], Pirozzoli and Bernardini [11][12]. The findings advance the knowledge front and enrich the contents of TBL theory and wall-bounded turbulence.

TBL scaling is a critically important issue in the traditional law [1] and the growing DNS data support to broadly classify generic TBL based on the distributions of time-averaged local wall-shear stress ($\tau_w$), a physical quantity directly associated with TBL scaling. Cao & Xu [13] pointed out the necessity of classification to re-understand the traditional law. The definition reads: Type-A with $\tau_w = \tau_w(x)$, Type-B with $\tau_w = \tau_w(z)$ or $\tau_w = \tau_w(y)$ and Type-C with $\tau_w = \tau_w(x,z)$ or $\tau_w = \tau_w(x,y)$, where $x$ is the streamwise direction, $y$ or $z$ is the wall normal or span-wise direction, respectively. A typical case for Type-A TBL is the semi-infinite flat-plate turbulence focused by the current paper. On the other hand, Type-B TBL can be represented by the turbulence in infinite long ducts driven by pressure gradient in streamwise direction, as seen in Cao & Xu [13] and Type-C TBLs are wall-bounded turbulence with more complex wall-structures which are featured by the local wall shear stress $\tau_w$ being the function of both streamwise and spanwise directions. With the classification, the traditional law was well-known validated for Type-A TBL and the law's applicability to Type-B TBL was preliminarily investigated in [13]. The paper further develops the complete law's formulations for Type-A TBL using both time-averaged local and ensemble-averaged scales and extends the law's applicability to thermal TBL. Although the Type-A TBL of semi-infinite plat-plate turbulence is geometrically simple, the investigations of the traditional base-line configuration, indeed, revealed an abundance of turbulent characteristics such as the hairpin vortices, coherent structures and recently vortex forest in the TBL. Fig.1 schematically displays the TBL patterns for semi-infinite plat-plate turbulence evolving along the streamwise $x$ direction while a variety of sublayers are formed in the course of turbulence evolution in the wall-normal y direction. The momentum exchange interactions are found promoted by the ejection and sweeping processes which eventually evolve into the formation of multi-eddies structures, namely the $\Lambda$-shaped vortices to hairpin vortices and further to vortex-forest structures, seen in FIG.1 (b) [14]. The higher momentum fluids are continuously sent into viscous inner layer by sweeping process resulting in the growing of wall shear stress and the development of TBL thickness of TBL along the streamwise direction.

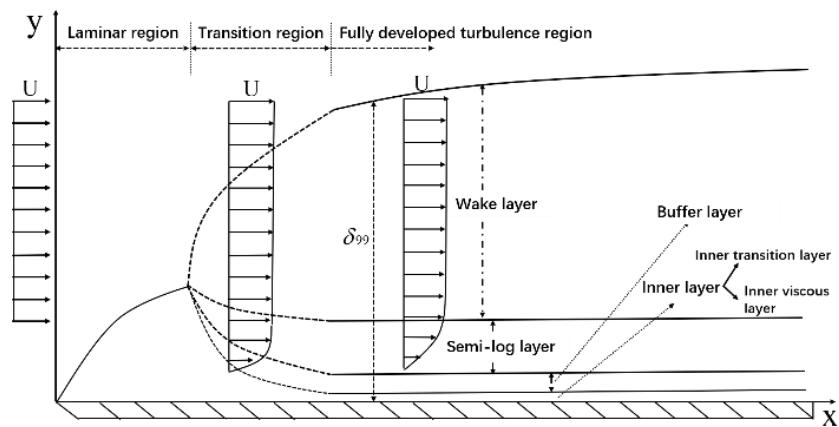

(a)

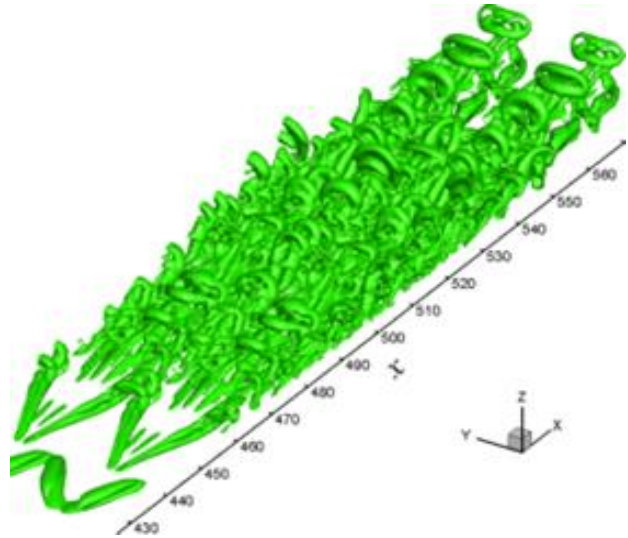

(b)

FIG.1: (a) Schematic drawing and (b) natural transition, Liu [14], of semi-infinite flat-plate turbulent flow.

According to [1], TBL velocity and length scales were identified by nondimensionalizing the definition of wall-shear stress ($\tau_w$), including the steps in Eq. (1) by making the left-hand side of definition equal to unity and reorganizing the right-hand side by nondimensionalizing the nominator and denominator in the partial derivative leading to the appropriate velocity and length scales.

$$\tau_w = \mu \frac{\partial u}{\partial y} \rightarrow 1 = \mu \frac{\partial u/\tau_w}{\partial y} = \mu \frac{\partial u/u_\tau}{\partial \rho u_\tau y/\mu} = \frac{\partial u^*}{\partial y^*} \tag{1}$$

Thereby, the TBL velocity and length scales are well-known defined as $u_\tau = \sqrt{\tau_w/\rho}$, $l_\tau = \mu/(\rho u_\tau)$, $y^* = y/l_\tau$, $u^* = u/u_\tau$, where $\rho$ is density, $\mu$ is dynamic viscosity. $Re$-independent model equations for Type-A TBL were then derived based on these scales in the time-averaged local sense [1].

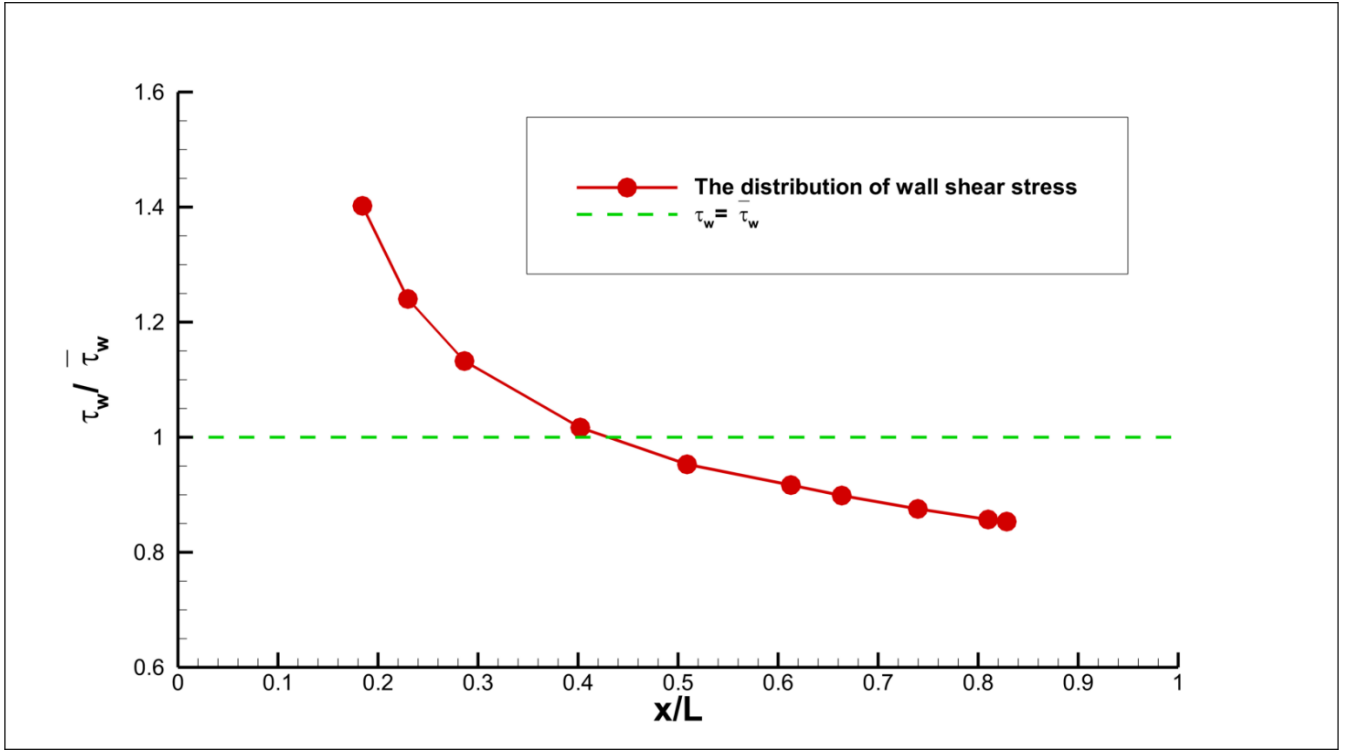

FIG.2: The typical wall shear stress distribution along flat plate surface [8][9] where the length scale $L$ is the total length of flat plate and stress scale $\overline{\tau}_w$ is the ensemble-averaged wall shear stress.

An important concept was introduced in [13] to distinguish $\tau_w$ in the traditional law from the ensemble-averaged wall shear stress ($\overline{\tau}_w$) in exploring the law for Type-B TBL from a new perspective. The approach can be directly applied to Type-A TBL by taking the space-averaging in $x$ direction on the time-averaged quantities. FIG.2 presents the typical distribution of wall shear stress along the flat-plate surface in streamwise direction [8][9]. Before the turbulent transition region, the wall shear stress takes a larger value, and then decreases with the increasing of TBL thickness. With the further evolution of wall turbulence, the wall shear stress slows down its decreasing rate. The ensemble-averaged wall shear stress $\overline{\tau}_w$ can be calculated as $\overline{\tau}_w = \int_{l1}^{l2} \tau_w(x)dx/(l_2 - l_1)$ where $l_1, l_2$ denote the starting and ending locations of the fully-developed TBL region on the plate under investigation. Then the ensemble-averaged scales of velocity $\overline{u}_\tau = \sqrt{\overline{\tau}_w/\rho}$ and length $\overline{l}_\tau = \mu/(\rho \overline{u}_\tau)$ can be determined. Since the ensemble-averaged quantities are no longer a function of independent variable $x^+ = \rho \overline{u}_\tau x/\mu$, the non-dimensional form of time-averaged governing equations for Type-A TBL can be written as

$$\frac{\partial \overline{u^+ u^+}}{\partial x^+} + \frac{\partial \overline{u^+ v^+}}{\partial y^+} = \frac{\partial^2 \overline{u^+}}{\partial x^{+2}} + \frac{\partial^2 \overline{u^+}}{\partial y^{+2}} + \frac{\partial \overline{u^{+\prime} u^{+\prime}}}{\partial x^+} + \frac{\partial \overline{u^{+\prime} v^{+\prime}}}{\partial y^+} \qquad (2)$$

$$\frac{\partial \overline{u^+ v^+}}{\partial x^+} + \frac{\partial \overline{v^+ v^+}}{\partial y^+} = -\frac{\partial \overline{p}^+}{\partial y^+} + \frac{\partial^2 \overline{v^+}}{\partial x^{+2}} + \frac{\partial^2 \overline{v^+}}{\partial y^{+2}} + \frac{\partial \overline{u^{+\prime} v^{+\prime}}}{\partial x^+} + \frac{\partial \overline{v^{+\prime} v^{+\prime}}}{\partial y^+} \qquad (3)$$

where $\overline{u}^+ = u/\overline{u}_\tau$, $x^+ = \rho \overline{u}_\tau x/\mu$, $y^+ = \rho \overline{u}_\tau y/\mu$.

Regarding to the thermal TBL, under the ensemble-averaged scales defined by the time-averaged wall-heat flux $\overline{q}_w$, the length and temperature scales can be defined as $y_t^+ = c\rho \overline{u}_\tau y/k = y^+/Pr$, $T^+ = T/\overline{T}_\tau$, where $\overline{T}_\tau = \overline{q}_w/(c\rho \overline{u}_\tau)$, $k$ is thermal conductivity coefficient, $c$ is specific heat and $Pr = k/(c\mu)$ is Prandtl number. The Fourier's law can then be written as $\partial T^+/\partial y_t^+ = q_w/\overline{q}_w$ and the governing equation as

$$\frac{\partial \overline{u^+ T^+}}{\partial x_t^+} + \frac{\partial \overline{v^+ T^+}}{\partial y_t^+} = \frac{\partial^2 \overline{T^+}}{\partial x_t^{+2}} + \frac{\partial^2 \overline{T^+}}{\partial y_t^{+2}} + \frac{\partial \overline{u^{+\prime} T^{+\prime}}}{\partial x_t^+} + \frac{\partial \overline{v^{+\prime} T^{+\prime}}}{\partial y_t^+} \qquad (4)$$

Obviously, the solution conditions for Eq.(2) (3) and (4) do not rely on any criterion numbers and the analytic expression of $u^+(x^+, y^+)$ based on Eq.(2) (3) is $Re_\tau$-independent and $T^+(x_t^+, y_t^+)$ in Eq.(4) is both $Re_\tau$ and $Pr$ independent. Therefore, based on the ensemble-averaged scales, the instantaneous DNS data can be rescaled and be independent on $Re_\tau$ and $Pr$. Logically, as the approximation of the numerical solution, the analytic formulations, specifically the control parameters designed based on GDF and GEF, are independent on $Re_\tau$ and $Pr$. These governing equations are the foundation to understand, prove and validate the complete analytical formulations of the unified law for both flow and thermal Type-A TBLs.

With the definitions of time-averaged local and ensemble-averaged scales, i.e. $(u_\tau, l_\tau)$ and $(\overline{u}_\tau, \overline{l}_\tau)$, and the introduction of wall-shear stress increment $\Delta^+(x) = \tau_w/\overline{\tau}_w - 1$, the traditional viscous linear law can be expressed in two equivalent forms, namely $u^+ = (1+\Delta^+)y^+$ in $(\overline{u}_\tau, \overline{l}_\tau)$ scales and $u^* = y^*$ in $(u_\tau, l_\tau)$ scales. $(\overline{u}_\tau, \overline{l}_\tau)$ can be converted to $(u_\tau, l_\tau)$ by $(u_\tau, l_\tau) = (\sqrt{1+\Delta^+}\,\overline{u}_\tau, 1/\sqrt{1+\Delta^+}\,\overline{l}_\tau)$. Hence, the traditional viscous linear law can be interpreted as a single-control parameter form, whereas the current work targets at the relevant law for each sublayer in a multi-control parameter form with improved accuracy.

Typical TBL contains multi-sublayers with different physical and mathematical mechanisms. Following the traditional TBL descriptions in Coles [15], the valid range for each sublayer has to be precisely defined so that the law expression can be accurately developed. However, the precise determination for each sublayer boundary has long been a tricky puzzle. As is well known, the TBL outer boundary was defined as the location of 99% the incoming velocity proposed by Prantdl [2]. To the author's point of view, the validity of the definition is based on the fact that, in the vicinity of TBL outer layer, the velocity gradient is trivial, and therefore the velocity profile is reasonably chosen to determine the boundary between TBL and outer inviscid flow. However, both velocity gradient and profile are equally important in the vicinity of wall, which suggests that these two quantities, i.e. profile and gradient, need to be taken into account to define the boundary of each TBL sublayer near wall.

Within the context, the paper proposed to utilize the indicator function (IDF) [10][12][16], with the form of combination, or specifically product of viscous linear velocity profile and gradient, i.e. $\mathrm{IDF}(y^*) = y^* \partial u^*/\partial y^* = \partial u^*/\partial \ln(y^*)$, to precisely define and quantitatively determine the boundaries of near-wall sublayers and the associated control parameters in corresponding law expressions, such as the von Karman constant $\kappa^*$ in traditional log-law and the $(\varepsilon_i^*, D_i^*)$, $i = 1, 2, 3$ in current study. Usually, the near-wall sublayers, such as the inner, buffer layers and the lower boundary of semi-log layer, were measured by the TBL length scale of $l_\tau$ or $\overline{l_\tau}$ and the sublayers far away from wall, such as the upper boundary of semi-log layer and wake layer, were determined by the percentage of TBL thickness $\delta_{99}^*(x)$ given as the location of 99% freestream velocity [2][15]. As depicted in FIG.3, the sublayer partition points were represented by $d_{sm}^*$ under the time-averaged local scales with subscript $m = 1,2,3,4,5$ denoting each sublayer's lower and upper boundaries. The paper gives the precise definitions for the partitions based on IDF and develops the corresponding law in each sublayer.

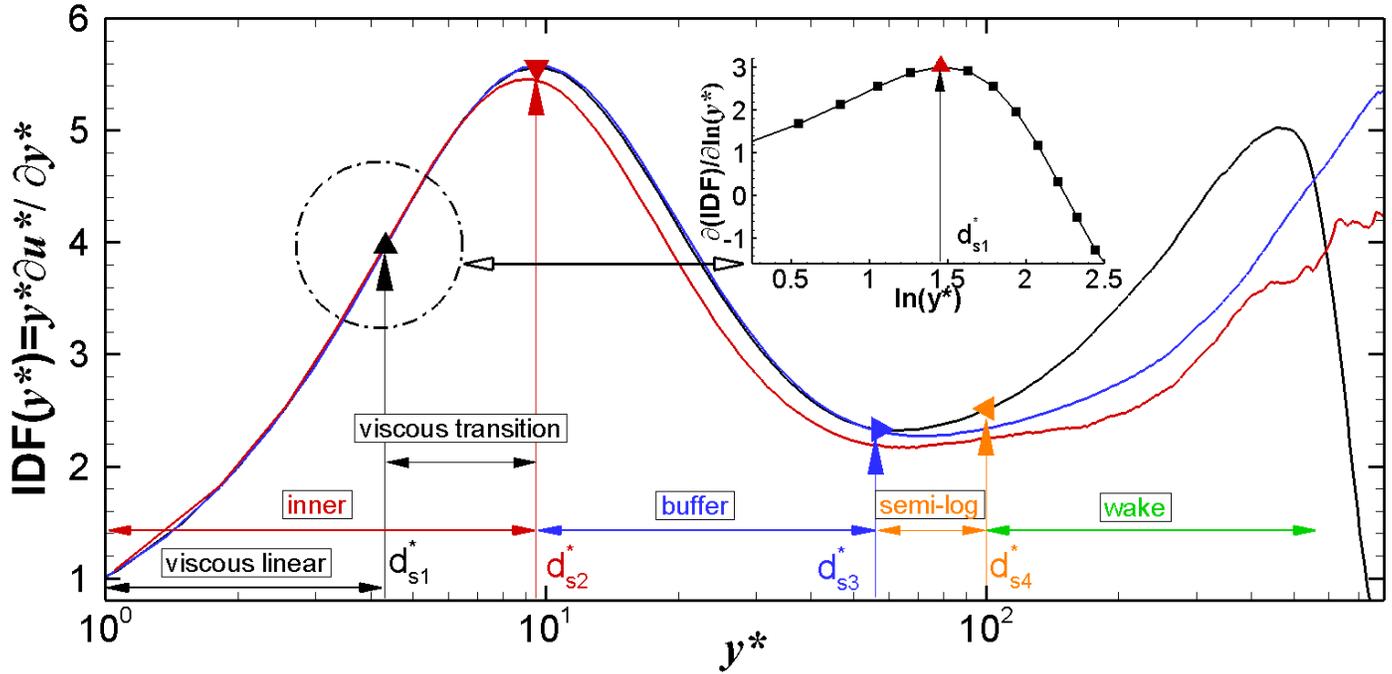

FIG.3: IDFs under semi-log coordinate with $Re_\theta = 2000, 4060$ from [8][9] (black, blue) and $Re_\theta = 6044, Mach = 2$ from [11][12] (red).

Obviously, the IDF is capable of depicting the velocity characters of each sublayer. The profiles in viscous linear and semi-log zones are featured by the linear and constant IDF, respectively. The buffer-layer IDF varies from a linear to a log-constant function. Hence, based on the IDF, the partitions of viscous linear, inner, buffer and semi-log layers can be definitely determined. FIG.3 illustrates the IDF with the characteristics in each sublayer under semi-log coordinate.

The velocity profiles grow linearly from wall resulting in a linear IDF with $\partial \text{IDF}(y^*)/\partial \ln(y^*) = y^*$ increasing linearly. However, with the rising of Reynolds stress away from wall, the profiles deviate from the linear law ($u^* = y^*$) and $\partial \text{IDF}(y^*)/\partial \ln(y^*)$ in FIG.3 slows down its growing until reaching a local maximum where the upper boundary of viscous linear layer $d_{s1}^*$ is defined, i.e. $\partial^2 \text{IDF}(y^*)/\partial \ln(y^*)^2 |_{d_{s1}^*} = 0$. As shown in FIG.3, $d_{s1}^*$ is the first characteristic location as an IDF's inflection point where the viscous linear law ends and the nonlinearity due to Reynolds shear stress comes into effectiveness. Hence $0 \leq d^* \leq d_{s1}^*$ can be rigorously determined as the region for viscous linear layer. The existing DNS data consistently support $d_{s1}^* \approx 4.2$ with a maximum relative velocity error of 2%. Comparing to the linear law, the velocity gradient relative error is within 5%, which indicates that the non-linearity, or particularly the Reynolds shear stress $\overline{u^{*'}v^{*'}}$, can be neglected within the viscous linear layer and beyond the IDF inflection point, the nonlinearity has to be considered. FIG.3 also suggests that the conclusion be valid not only for incompressible, but also for compressible turbulence.

The semi-log law is featured by the reciprocal of von Karman constant IDF $(1/\kappa^*)$ with $\partial \mathrm{IDF}(y^*)/\partial \ln(y^*) = 0$, which implies that an IDF peak exist in between viscous linear and semi-log layers. The location is the second characteristic point on the IDF curve, denoted as $d_{s2}^*$ in FIG.3 under $l_\tau$ scales. Between $d_{s1}^*$ and $d_{s2}^*$, neither viscous stress nor Reynolds shear stress can be neglected. However, the viscous shear stress $\tau^*(y^*)$ still plays a dominant role in shaping the velocity profiles over the Reynolds shear stress $\partial \overline{u^{*'}v^{*'}}/\partial y^*$, therefore the IDF maintains an increasing mode until $d_{s2}^*$. The layer in $d_{s1}^* \leq d^* \leq d_{s2}^*$ is traditionally a part of buffer layer, but currently a viscous transition layer as part of the inner layer within which the profiles can be expressed as the viscous linear law corrected by the GDF or GEF [13]. The definition of $d_{s1}^*$ and $d_{s2}^*$ can be directly applied to define the upper boundary of inner viscous linear sublayer $d_{s1}^+$ and inner viscous transition sublayer $d_{s2}^+$ under the ensemble-averaged scale $(\overline{u_\tau}, \overline{l_\tau})$, resulting in the relation of $d_{si}^* = (\overline{l_\tau}/l_\tau)d_{si}^+ = d_{si}^+\sqrt{1+\Delta^+}$ where $i = 1,2$. In order to clearly observe the DNS results under the unified $(\overline{u_\tau}, \overline{l_\tau})$ and $(u_\tau, l_\tau)$ scales and to examine the characteristics of $\overline{u^{+'}v^{+'}}$ and $u_{rms}^{+'}$, FIG.4 presents the distributions of $\overline{u^{+'}v^{+'}}$ and $u_{rms}^{+'}$ (the root of mean square) which characterizes the turbulence in viscous linear and viscous transition layers. It is evident to see that, starting from the location of $y^+ = d_{s1}^+$ as defined by the inflection point of IDF, the Reynolds shear stress accelerates its dropping rate and the turbulent fluctuation slows down its increasing rate in the viscous transition layer. The same result also holds for $\overline{u^{*'}v^{*'}}$ under local time-averaged scale, because changing scales only amplify or shrink the non-dimensional physical quantities, but not their distribution characteristics. Moreover, the scales $(\overline{u_\tau}, \overline{l_\tau})$ and $(u_\tau, l_\tau)$ are almost equivalent when $\mathrm{Re}_\theta = 2000$ presented by blue lines in FIG.4. Therefore, the blue lines can also be regarded as the profile under $(u_\tau, l_\tau)$ scale.

If the lower boundary of semi-log layer is denoted by $d_{s3}^*$, the zone of $d_{s2}^* \leq d^* \leq d_{s3}^*$ is then a buffer layer with $d_{s3}^*$ being the location where the IDF drops to a constant, i.e. $(1/\kappa^*)$, and the shaping effect of Reynolds shear stress on velocity profiles becomes stronger than the viscous shear stress. The profiles within the layer transit from the inner-layer to semi-log law. The satisfaction of $\partial \mathrm{IDF}(y^*)/\partial \ln(y^*)|_{d_{s2}^*} = \partial^2 u^*/\partial \ln(y^*)^2|_{d_{s2}^*} = 0$ indicates that $d_{s2}^*$ is an inflection point on the velocity profiles under semi-log coordinate. Similar to the analysis of Reynolds shear stress, turbulent fluctuation is also plotted under the scale $(\overline{u_\tau}, \overline{l_\tau})$ and $(u_\tau, l_\tau)$. Fig.4 further demonstrates that $d_{s2}^+$ or $d_{s2}^*$ is also an inflection point on the curve of Reynolds shear stress and after $d_{s2}^+$ or $d_{s2}^*$, the turbulent fluctuation $u_{rms}^+$ or $u_{rms}^*$ approaches to maximum and then decrease with a mild slope. The paper found that the profiles in buffer layer were able to be analytically expressed by the GDF and GEF. Since IDF is more directly linked to the velocity gradient or total shear stress, it is more appropriate to use the shear force related IDF instead of the velocity profiles to define boundaries of each sublayers, particularly for the inner layer where the velocity gradient is large, which is considered more consistent with the physical features of inner and buffer sublayers.

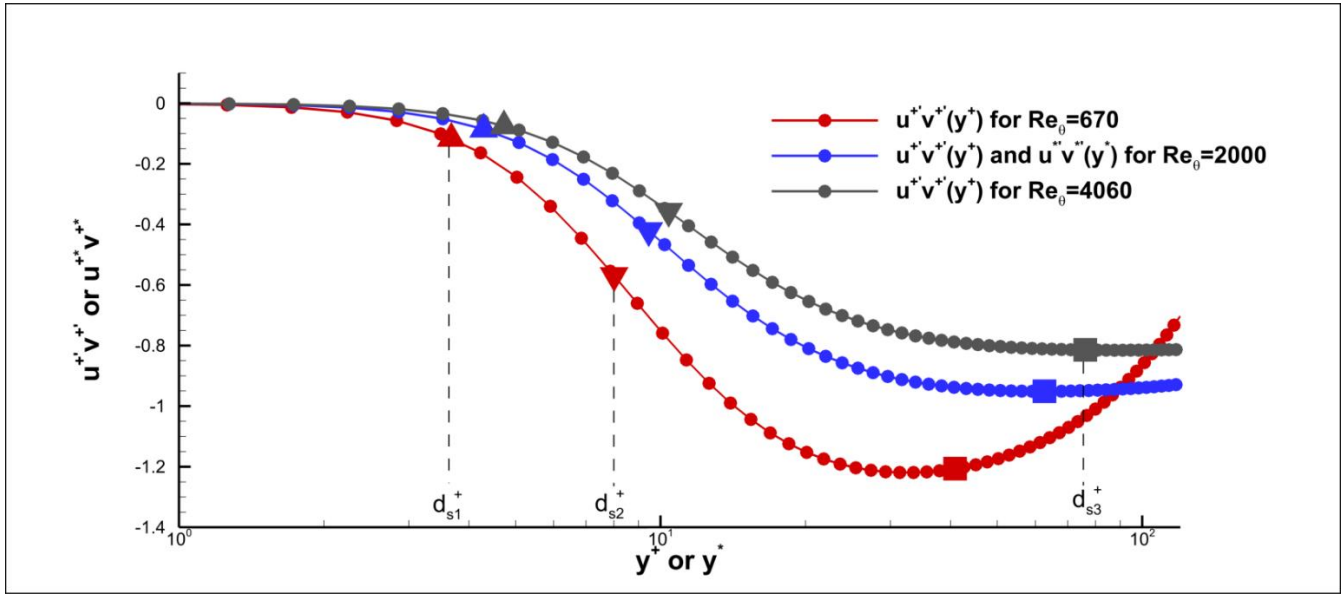

(a)

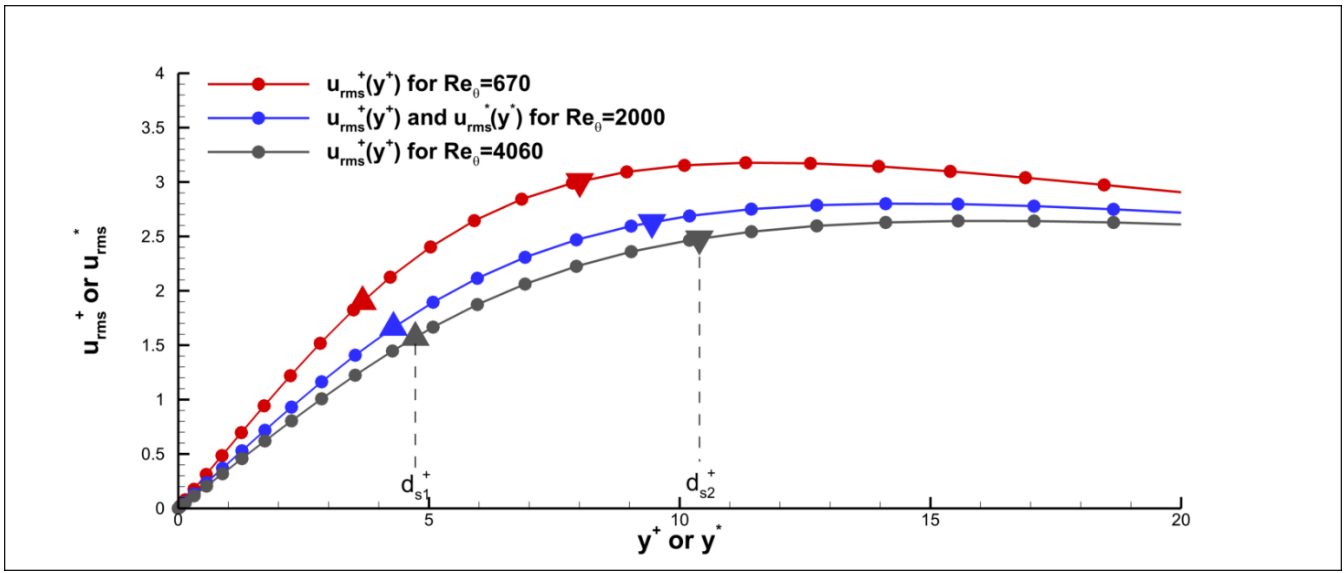

(b)

FIG.4: the distributions of (a) Reynolds shear stress $\overline{u^{+'}v^{+'}}$ and $\overline{u^{*'}v^{*'}}$ from and (b) turbulent fluctuations $u_{rms}^{+}$ and $u_{rms}^{*}$ for Type-A TBL with different Reynolds numbers [8] where delta, gradient and square denote $d_{s1}^{+}$ or $d_{s1}^{*}$, $d_{s2}^{+}$ or $d_{s2}^{*}$ and $d_{s3}^{+}$ or $d_{s3}^{*}$.

In the region adjacent to semi-log layer, the IDF reaches the minimum and tends to be a constant for smaller $Re$ shown in FIG.3, or increases slightly after reaching the minimum and then approaches to a constant for larger $Re$ [10][12]. The IDF minimum is slightly smaller than $1/\kappa^{*}$, and the location can be used to defined $d_{s3}^{*}$. By denoting $d_{m}^{*}$ as the minimum IDF location, $d_{s3}^*$ can be determined as $0.9d_m^*$ for relatively small $Re$ found in the paper and $1.2 \sim 2.0 d_m^*$ for larger $Re$ given in [17]. The DNS data used in the paper had relatively low $Re$ with $d_{s3}^*$ varying from 50 to 70, slightly larger than the conventional 50 or even 30 in [11][18]. The Reynolds shear stress maintains nearly a constant before $d_{s3}^*$ for relatively large Reynolds numbers, resulting in an obvious semi-log sublayer which is considered as a distinctive feature for semi-log layer and is consistent with the traditional theory based on the Prandtl mixing length theory [2] (see FIG.3 or 4(a)).

The transition point from semi-log to wake layer is denoted by $d_{s4}^*$ and the location was suggested being measured by $\delta_{99}^*$ with an averaged factor of $0.15$ [19] proposed, resulting in $d_{s4}^* = 0.15\delta_{99}^*$ and $d_{s5}^* = \delta_{99}^*$. Therefore, based on the above discussions, the sublayers are defined precisely in $d_{s(m-1)}^\# < d^\# \leq d_{sm}^\#$ with $d_{s0}^\#=0$, $\#=*$ for $(u_\tau, l_\tau)$ and $\#=+$ for $(\overline{u_\tau}, \overline{l_\tau})$ scales, $m=1$ being the viscous linear, $m=2$ the inner transition, $m=3$ the buffer, $m=4$ the semi-log and $m=5$ the wake layers. Moreover, the inner layer is formed by merging viscous linear and inner transition layers.

Starting from von Karman [1], the explorations of TBL law were mainly based on solving the simplified $Re$-independent RANS model with the Reynolds shear stress being treated by various hypotheses or physically-guided fittings under the $(u_\tau, l_\tau)$ scaling [12][20][21]. The current paper extends the law exploration to the $(\overline{u_\tau}, \overline{l_\tau})$ scaling and establishes the $Re$-independent RANS equations, which permits more rigorous analysis using DNS data.

Another important issue is that the law's mathematical forms have to be consistent under different scales, namely the $(u_\tau, l_\tau)$ and $(\overline{u_\tau}, \overline{l_\tau})$ scales, and such forms ought to be well-posed to maintain the basic mathematics universally applicable to relevant scales by only adjusting the control parameters. Evidently, the issue is satisfactorily resolved in the following analysis by identifying the proper scale quantity as $u_\tau/\overline{u_\tau} = \sqrt{1+\Delta^+}$ and by rescaling the other control parameters using $\sqrt{1+\Delta^+}$.

With precisely defining the boundaries of sublayers, the paper then proceeds to the analytical design for the law in each sublayer. For convenience, the following discussion of formulation design is under the local time-averaged scale $(u_\tau, l_\tau)$ and the same method can be applied using the ensemble-averaged scale $(\overline{u_\tau}, \overline{l_\tau})$. In $0 \leq d^* \leq d_{s1}^*$, the dominant term is wall-shear stress and the velocity profiles were traditionally expressed by the exponential function under semi-log coordinate, i.e. the viscous linear law $u^* = e^{\ln(y^*)}$. In $d_{s1}^* \leq d^* \leq d_{s2}^*$, the emerging Reynolds shear stress makes the velocity profiles be suppressed below $u^* = e^{\ln(y^*)}$, implying a damping factor be introduced to reflect the velocity deficit. Inspired by the van Driest function [22] and the GDF [13], an exponential function is proposed as the damping factor. Meanwhile, the well-known linear and semi-log linear laws suggest that a mixed-scale phenomenon occurs and a mixed-law in between be logically designed for the inner transition layer. Hence the damping mechanism takes the form of exponential function with a linear combination of $\ln(y^*)$ and $y^*$ to represent the transition from a linear to logarithmic character. The law expression in the inner layer can then be designed by the form of Eq. (5). It is obvious that the inner-layer law Eq. (5) was constructed based on the traditional linear law and the newly introduced GDF and GEF. Similarly, the analytical expression of velocity profiles in the buffer and wake layers can also be logically designed by the semi-log linear law and the in-depth understandings of GDF and GEF. Moreover, the traditional linear and semi-log laws can be interpreted as an identical mathematics in the form of linear function with the independent variable being expressed as the linear coordinate $y^*$ or the logarithmic coordinate $\ln(y^*)$. The velocity profiles in the buffer and wake layers can thus be analytically expressed in the same mathematical form of Eq. (5) by switching the linear coordinate $y^*$ to the semi-log coordinate $\ln(y^*)$. Therefore, with each sublayer's partition being precisely defined by IDF and the GDF and GEF being applied to both linear and semi-log coordinates, the complete law formulae for entire TBL are derived by shifting the coordinate origin to the boundary locations of semi-log layer. These law formulations can then be uniformly expressed with the superscript #=(*,+) representing the $(u_\tau, l_\tau)$ and $(\overline{u_\tau}, \overline{l_\tau})$ scales, respectively, and $\Delta^* \equiv 0$, $\Delta^+ = \tau_w/\overline{\tau_w} - 1$.

(1) inner layer

$$u^\#(y^\#) = [1 + \Delta^\#(x^\#) + \varepsilon_1^\#(x^\#)]e^{\ln y^\#} - \varepsilon_1^\#(x^\#)e^{\ln y^\# + y^\#/D_1^\#(x^\#)} = y^\#[1 + \Delta^\#(x^\#) + \varepsilon_1^\#(x^\#) - \varepsilon_1^\#(x^\#)e^{y^\#/D_1^\#(x^\#)}] \tag{5}$$

where $0 \leq y^\# < d_{s2}^\#$, $\varepsilon_1^\#(x^\#)$ represents the damping strength, $D_1^\#(x^\#)$ stands for the weight between the linear and log scales and $\varepsilon_1^\#(x^\#)y^\#$ is included to satisfy the boundary condition of wall-shear stress;

(2) buffer layer

$$u^\#(y^\#) = u^\#(d_{s3}^\#) + \ln(y^\#/d_{s3}^\#)[1/\kappa^\# - \varepsilon_2^\#(x^\#) + \varepsilon_2^\#(x^\#)e^{-\ln(y^\#/d_{s3}^\#)/D_2^\#(x^\#)}], d_{s2}^\# \leq y^\# < d_{s3}^\# \tag{6}$$

(3) semi-log layer

$$u^\#(y^\#) = 1/\kappa^\# \ln(y^\#) + C^\#, d_{s3}^\# \leq y^\# < d_{s4}^\# \tag{7}$$

(4) wake layer

$$u^\#(y^\#) = u^\#(d_{s4}^\#) + \ln(y^\#/d_{s4}^\#)[1/\kappa^\# + \varepsilon_3^\#(x^\#) - \varepsilon_3^\#(x^\#)e^{-\ln(y^\#/d_{s4}^\#)/D_3^\#(x^\#)}], d_{s4}^\# \leq y^\# < d_{s5}^\# \tag{8}$$

where $(\kappa^*, C^*)$ are standard von Karman and wall-roughness constants. Similarly to the traditional linear and semi-log linear laws, the expression Eq. (5)(6)(7)(8) maintain the identical mathematics under different scales and the control parameters in Eq. (5)(6)(7) and (8), including $(\kappa^\#, C^\#)$ and $(\varepsilon_i^\#, D_i^\#), i = 1, 2, 3$, can be converted from $(\overline{u_\tau}, \overline{l_\tau})$ to $(u_\tau, l_\tau)$ scales by the relations $\varepsilon_i^+ = (1 + \Delta^+)\varepsilon_i^*, i = 1, 2, 3$. $\kappa^+ = \kappa^*/\sqrt{1+\Delta^+}$, $D_1^+ = D_1^*/\sqrt{1+\Delta^+}$, $D_2^+ = D_2^*$, $D_3^+ = D_3^*$ with $\Delta^+$ being the rescaling parameter. FIG.5 illustrates the damping effect of Eq. (5). The inner-layer velocity profiles with $Re_\theta = 1000$ and $Re_\theta = 3970$ lie above and below the neutral linear relation $u^+ = y^+$ under the unified ensemble-averaged scale $(y^+, u^+)$, which reflects the enhancing or damping effect of switching scales from $(y^*, u^*)$ to $(y^+, u^+)$. Before $d_{s1}^+$, the linear law $u^+ = (1 + \Delta^+)y^+$ (single-parameter form) and dual-control parameter formulation $u^+ = y^+(1 + \Delta^+ e^{y^+/D})$ porposed by Xu [13] and tri-parameter formulation of Eq. (5) coincide with each other. Morever, comapred with the linear law, both formulations of $u^+ = y^+(1 + \Delta^+ e^{y^+/D})$ and Eq. (5) are damped, which are closer to the DNS velocity profiles. Although, overall, the formulation $u^+ = y^+(1 + \Delta^+ e^{y^+/D})$ provides a more accurate inner-layer velocity fitting, it deviates from the velocity profiles and can not capture the velocity gradient in the vicinity of $d_{s2}^+$, as demonstrated in FIG.5 (a). Eq. (5) can not only descibe the velocity and its gradient precisely, but also smoothly connected to the buffer-layer velocity profiles and gradient at $d_{s2}^+$.

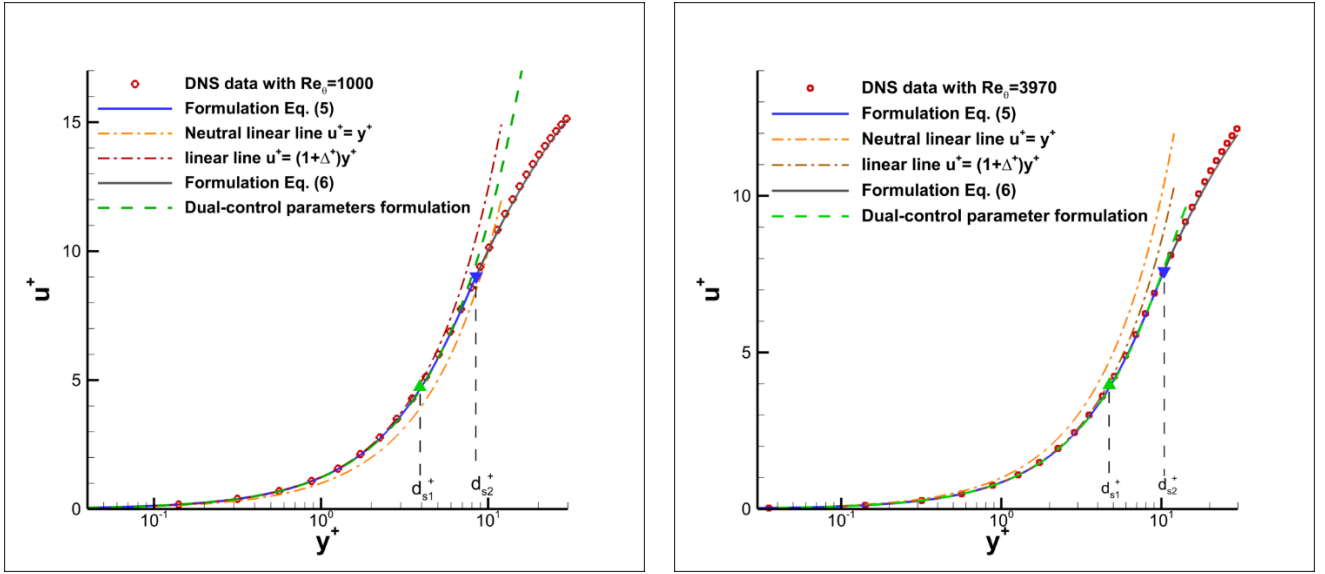

FIG.5: The inner layer formulation enhancing or damping effect compared with the tradition linear law for different Reynolds numbers; $Re_\theta = 1000$ in (a) and $Re_\theta = 3970$ in (b) [8][9] where green delta and blue gradient denote the location of $d_{s1}^+$ and $d_{s2}^+$.

The formulations Eq. (5)(6)(7)(8) can generally be considered as tri-control-parameter law, as compared to the single-control-parameter traditional law for viscous linear layer mentioned before and the dual-control-parameter law for the inner layer of Type-B TBL [13]. When using $(\overline{u_\tau}, \overline{l_\tau})$ as scales, the control parameters are all explicitly presented in Eq. (5)(6)(7) and (8) by $(\Delta^+, \varepsilon_i^+, D_i^+, i=1,2,3)$ and $(\Delta^+, \kappa^+, C^+)$ in the four sublayers. However, under $(u_\tau, l_\tau)$ scales, the variable $u_\tau$, or the equivalent $\sqrt{1+\Delta^+}\,\overline{u_\tau}$, is merged into the variables of $u^*$ and $(x^*, y^*)$, resulting in the law being uniformly expressed by the superscript $\#=(*,+)$ under $(u_\tau, l_\tau)$ and $(\overline{u_\tau}, \overline{l_\tau})$ scales, respectively. It is worth to note that the parameters $(\Delta^+, \varepsilon_i^+, D_i^+, i=1,2,3)$ and $(\Delta^+, \kappa^+, C^+)$ are $Re_\tau$-independent if Eq. (5)(6)(7)(8) are the approximate solutions for Eq. (2)(3), and therefore, in general, these parameters are universally applicable to any $Re_\tau$. In order to profoundly understand the control parameters $(\varepsilon_i^+, D_i^+, i=1,2,3)$ in Eq. (5)(6)(8), a relative deviation function of $RD$ comparing to the linear law can be written as:

$$RD_{linear} = [y^\#[1+\Delta^\# + \varepsilon_1^\# - \varepsilon_1^\# e^{y^\#/D_1^\#}] - y^\#(1+\Delta^\#)]/y^\#(1+\Delta^\#) = \frac{\varepsilon_1^\#}{1+\Delta^\#}(1-e^{y^\#/D_1^\#}) \qquad (9)$$

The term $\varepsilon_1^\#/(1+\Delta^\#)$ determines the damping strength and $D_1^\#$ represents a characteristic length reflecting the location where Reynolds shear stress start to affect velocity profiles. For a fixed damping strength $\varepsilon_1^\#$, the larger $D_1^\#$ implies that $y^\#$ need to take a larger value so that the term $(1-e^{y^{\#}/D_1^{\#}})$ is not significantly small and the velocity profiles start to deviate from the linear law. Similarly, $(\varepsilon_i^+, D_i^+, i=2,3)$ have the same mathematical meaning as $(\varepsilon_1^+, D_1^+)$ under the semi-log rather than linear coordinate. The mathematics of law are fundamentally identical for the scales of $(\overline{u_\tau}, \overline{l_\tau})$ and $(u_\tau, l_\tau)$, giving rise to the unified and consistent formulae with high fidelity to the physics in each sublayer, as demonstrated in following validation.

To validate Eq. (5)(6)(7)(8), the DNS data for Type-A TBLs were utilized to determine the relevant control parameters, to estimate the law's accuracy and to prove the $Re_\tau$ independency and similarity. It is important to note that the scaling control parameter $\Delta^+$ has to be precisely given by DNS data so that the scales of $(\overline{u_\tau}, \overline{l_\tau})$ and $(u_\tau, l_\tau)$ are switchable and the following validation can be conducted. The physical conditions satisfied at the partition points can then be applied to calculate the control parameters of $(\varepsilon_i^{\#}, D_i^{\#}, i=1,2,3)$.

For viscous linear layer, due to the TBL thin layer and large velocity gradient assumptions, the dependence of $(\varepsilon_1^{\#}, D_1^{\#})$ on $x^{\#}$ was omitted, the velocity $u^{\#}$ and shear stress $\tau^{\#} = \partial u^{\#}/\partial y^{\#}$ at $d_{s2}^{\#}$ from the DNS data were used to calculate $(\varepsilon_1^{\#}, D_1^{\#})$ by solving Eq. (10) obtained by Eq. (5) with a given velocity $u^{\#}(d_{s2}^{\#})$ and its gradient $\tau^{\#}(d_{s2}^{\#})$ at $d_{s2}^{\#}$ from a DNS solution

$$u^{\#}(d_{s2}^{\#}) = d_{s2}^{\#}(1+\Delta^{\#}+\varepsilon_1^{\#}-\varepsilon_1^{\#}e^{d_{s2}^{\#}/D_1^{\#}}), \quad \tau^{\#}(d_{s2}^{\#}) = 1+\Delta^{\#}+\varepsilon_1^{\#}-\varepsilon_1^{\#}(1+d_{s2}^{\#}/D_1^{\#})e^{d_{s2}^{\#}/D_1^{\#}} \qquad (10)$$

For semi-log layer, $d_{s3}^{\#}$ and $d_{s4}^{\#}$ denote the lower and upper boundaries. The von Karman constant $\kappa^{\#}$ is a well-known parameter controlling the derivative of semi-log law, with $\kappa^* \approx 0.41$ for moderate $Re$ and slightly smaller than $0.41$ for high $Re$. With the given $\kappa^*$ and $\kappa^+ = \kappa^*/\sqrt{1+\Delta^+}$, $(\varepsilon_2^+, D_2^+)$ can be computed similarly with Eq. (10) by solving Eq. (11),

$$u^{\#}(d_{s2}^{\#}) = u^{\#}(d_{s3}^{\#}) + \ln(d_{s2}^{\#}/d_{s3}^{\#})[1/\kappa^{\#} - \varepsilon_2^{\#} + \varepsilon_2^{\#}e^{-\ln(d_{s2}^{\#}/d_{s3}^{\#})/D_2^{\#}}], \quad \tau^{\#}(d_{s2}^{\#}) = 1/d_{s2}^{\#}\{1/\kappa^{\#} - \varepsilon_2^{\#} + \varepsilon_2^{\#}[1-\ln(d_{s2}^{\#}/d_{s3}^{\#})/D_2^{\#}]e^{-\ln(d_{s2}^{\#}/d_{s3}^{\#})/D_2^{\#}}\} \qquad (11)$$

In wake layer near mainstream where $\partial u^+/\partial y^+$ becomes insignificant, velocity is the major quantity to predict. The $(\varepsilon_3^{\#}, D_3^{\#})$ in Eq. (8) can be computed by the velocities at two typical points. Since the wake layer occupies a large portion of TBL, the locations at $0.7\delta_{99}^{\#}$ and $0.9\delta_{99}^{\#}$ are suggested for the parameter estimation to give a satisfactory accuracy. The parameters can be determined by solving Eq. (8) for $(\varepsilon_3^{\#}, D_3^{\#})$ at the locations. FIG.6 (a) and (b) provides the velocity from the DNS data comparing to the predictions by Eq. (8) at the typical $Re_\theta$ under $(\overline{u_\tau}, \overline{l_\tau})$ and $(u_\tau, l_\tau)$ scales.

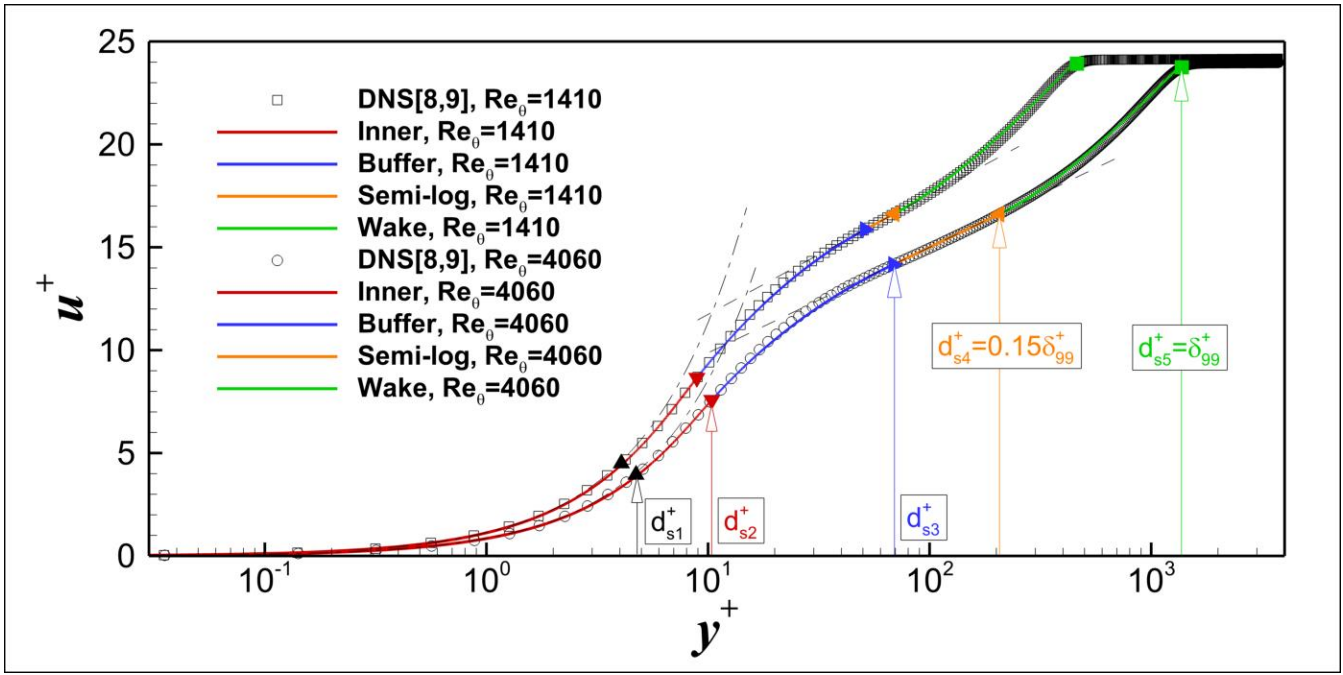

(a)

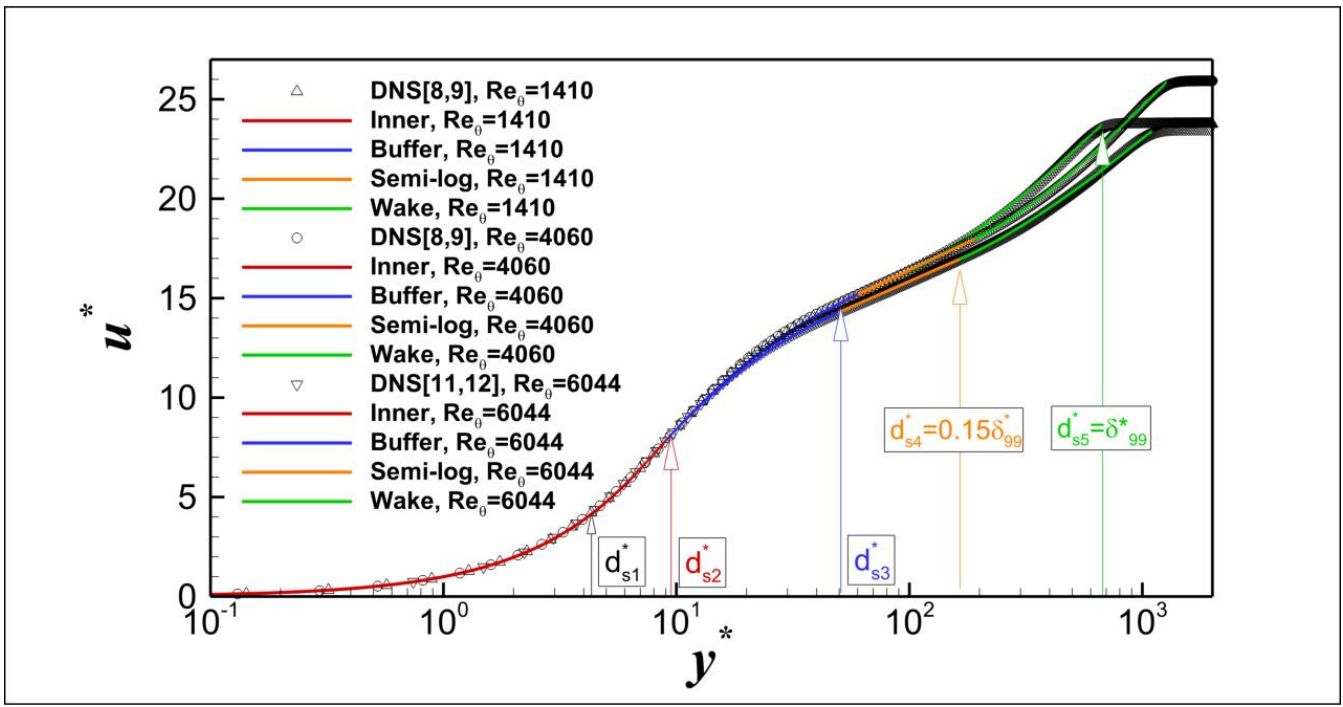

(b)

FIG.6: The velocity profiles under (a) $(\overline{u_\tau}, \overline{l_\tau})$ and (b) $(u_\tau, l_\tau)$ scales.

For quantitative validation, the relative error $Err_f$ of $f$ is defined by $Err_f = (f_{law} - f_{dns})/f_{dns}$ with $f_{law}$ and $f_{dns}$ being the quantity from the law prediction and DNS computation, respectively, and $f$ represents either velocity $u^\#$ or velocity gradient $\tau^\# = \partial u^\#/\partial y^\#$. FIG.6 indicates that the law formulation Eq. (5)(6)(7)(8) generally well agree with the DNS data in the entire TBL under $(\overline{u_\tau}, \overline{l_\tau})$ and $(u_\tau, l_\tau)$ scales. Quantitatively, the mean and maximum $Err_{u^\#}(mean, \max) \approx (0.01, 0.02)$ were found for the DNS data at $Re_\theta$ ranging from 670 to 6044. For the velocity gradients starting from the maximum $\tau_w^\# = (\partial u^\#/\partial y^\#)_{y^\#=0}$ at wall, $\tau^\#$ decreases remarkably near wall and tends to reduce to an insignificant amount $0.05\tau_w^\#$ around the mid of $d_{s2}^\# \le d^\# \le d_{s3}^\#$ shown in FIG. 6. Therefore, within the inner layer with $\partial u^\#/\partial y^\#$ always being significant, the mean and maximum $Err_{\tau^\#}$ were found at $Err_{\tau^\#}(mean, \max) \approx (0.02, 0.04)$. From the analysis of DNS data, $\tau^\# = \partial u^\#/\partial y^\#$ decreases remarkably to about $0.5\tau_w^\#$ at $d_{s2}^\#$ and further to about $0.05\tau_w^\#$ around the middle point of $d_{s2}^\# \le d^\# \le d_{s3}^\#$ which indicates $\tau^\# = \partial u^\#/\partial y^\#$ reaches a relatively small value at the outer boundary of inner layer. Beyond the buffer layer, the shear stress becomes insignificant, and therefore the quantitative validation for shear stress were conducted up to the buffer layer. More specifically, Table.1 lists the mean and max relative errors of the law's formulations, i.e. $Err_{u^\#}(mean, \max)$ of Eq. (5)(6)(8) in each sublayer compared with DNS data. The momentum Reynolds number and Mach number are denoted as $Re_\theta$ and $M\infty$ where $M\infty=/$ represents for the cases of incompressible wall-bounded turbulence.

Table.1: Relative errors of expressions Eq. (5)(6)(8) for different cases of $Re_\theta$ =1410, $M\infty$=/ for TBL1 [8][9], $Re_\theta$ =2000, $M\infty$=/ for TBL2 [8][9], $Re_\theta$ =4060, $M\infty$=/ for TBL3 [8][9], $Re_\theta$ =2241, $M\infty$=/ for TBL4 [10], $Re_\theta$ =8183, $M\infty$=/ for TBL5 [10], $Re_\theta$ =2866, $M\infty$=2 for TBL6 [11][12] and $Re_\theta$ =6044, $M\infty$=2 for TBL7 [11][12].

| Dataset | Inner layer $Err_{u^\#}(mean)$ | Inner layer $Err_{u^\#}(\max)$ | Inner layer $Err_{\tau^\#}(mean)$ | Inner layer $Err_{\tau^\#}(\max)$ | Buffer layer $Err_{u^\#}(mean)$ | Buffer layer $Err_{u^\#}(\max)$ | Wake layer $Err_{u^\#}(mean)$ | Wake layer $Err_{u^\#}(\max)$ |
|---|---|---|---|---|---|---|---|---|
| TBL1 | 0.0107 | 0.0194 | 0.0220 | 0.0437 | 0.0052 | 0.0121 | 0.0047 | 0.0101 |
| TBL2 | 0.0119 | 0.0202 | 0.0218 | 0.0420 | 0.0059 | 0.0131 | 0.0038 | 0.0084 |
| TBL3 | 0.0111 | 0.0202 | 0.0223 | 0.0437 | 0.0074 | 0.0157 | 0.0034 | 0.0076 |
| TBL4 | 0.0118 | 0.0212 | 0.0219 | 0.0418 | 0.0057 | 0.0129 | 0.0038 | 0.0084 |
| TBL5 | 0.0121 | 0.0220 | 0.0227 | 0.0429 | 0.0080 | 0.0171 | 0.0026 | 0.0069 |
| TBL6 | 0.0103 | 0.0170 | 0.0212 | 0.0399 | 0.0049 | 0.0112 | 0.0021 | 0.0063 |
| TBL7 | 0.0106 | 0.0180 | 0.0223 | 0.0421 | 0.0063 | 0.0139 | 0.0026 | 0.0058 |

Overall, the law formulations proved their capability of not only accurately predicting the profiles, but also capturing the large velocity gradients or shear stress behavior in inner layer. Moreover, the continuities of both $u^\#(y^\#)$ and $\partial u^\#/\partial y^\#$ are guaranteed at the partition points.

To make use of Eq. (5)(6) in application, the velocity and its gradient at $d_{s2}^\#$ have to be determined precisely. The paper obtained these quantities from the DNS data and found the velocity profiles were self-similar in $0 \le d^* \le d_{s2}^*$ as shown in FIG.6 (b) for the zero-pressure-gradient flat-plate turbulent TBL. Therefore, the velocity in inner layer can be represented by the law with parameters universally determined by the DNS data as $\varepsilon_1^*=0.0493$, $D_1^*=6.96$, $d_{s2}^*=9.605$, $u^*(d_{s2}^*)=8.1965$, $\tau^*(d_{s2}^*)=0.5831$ under $(u_\tau, l_\tau)$ scales. Moreover, the parameters $\kappa^\#, C^\#$ in semi-log layer are determined by the velocities at $d_{s3}^\#$ and $d_{s4}^\#$. Wake-layer parameters $\varepsilon_3^\#, D_3^\#$ are proposed to be calculated by the velocities at $0.7\delta_{99}^\#$ and $0.9\delta_{99}^\#$. Since $d_{s3}^*$ depends on $\partial u^*/\partial y^*$ based on IDF, $d_{s3}^*$ is suggested to be straightforwardly estimated as $d_{s3}^*=60$. All the relevant parameters in $(\overline{u_\tau}, \overline{l_\tau})$ scales can thus be obtained by the rescaling relations defined earlier.

The parameters in semi-log and wake layers are relatively insensitive to their locations, therefore the above methods are robust to determine the control parameters and to generate TBL profiles. FIG.7 presents the TBL in $(\overline{u_\tau}, \overline{l_\tau})$ scales with relatively high $Re_\theta$ from the experiment data [17]. The corresponding predictions from law with the parameters in Eq. (6)(7)(8) were calculated using the velocities at the four points of $d_{s3}^+, d_{s4}^+, 0.7\delta_{99}^+, 0.9\delta_{99}^+$. The law Eq. (6)(7)(8) are in good agreement with the experiments.

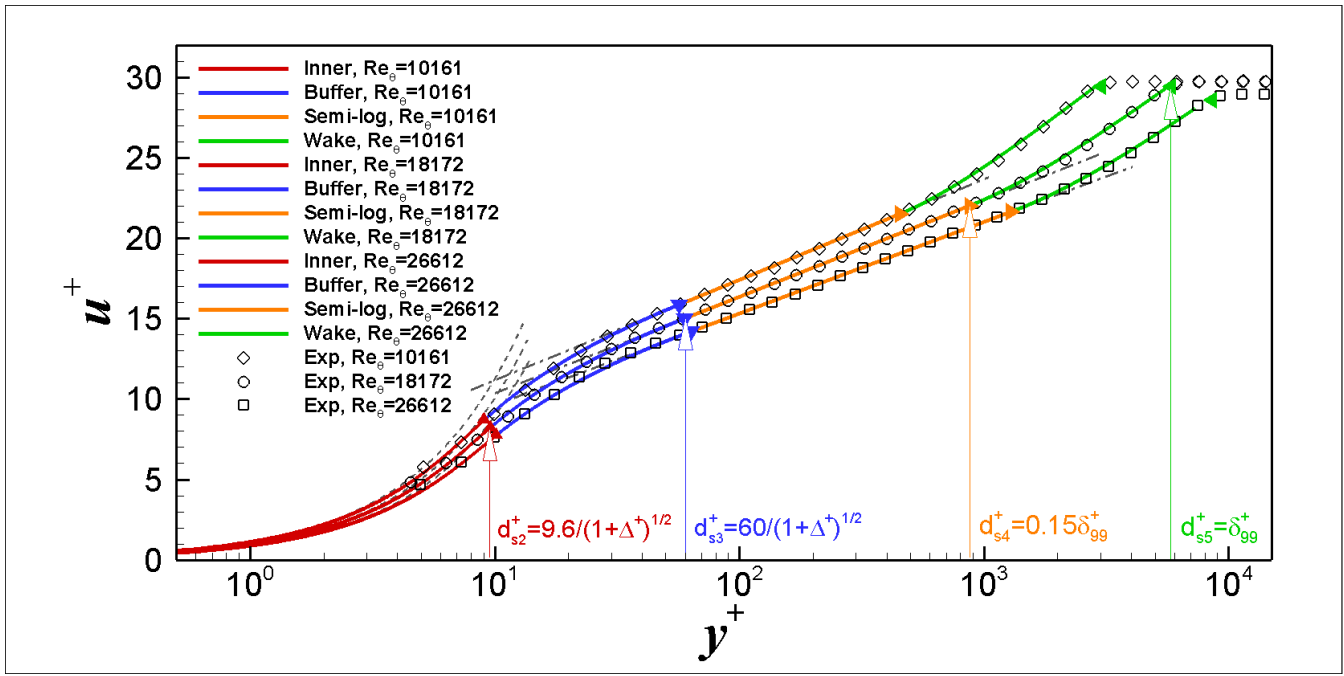

FIG.7: Type-A TBL law comparing with the experiment (Exp) data [17] under $(\overline{u_\tau}, \overline{l_\tau})$ scales.

In conclusion, the complete law-of-the-wall was investigated thoroughly under $(\overline{u_\tau}, \overline{l_\tau})$ and $(u_\tau, l_\tau)$ scales for Type-A TBL. The paper partitioned the TBL into four sublayers based on the IDF. Based on the partition, the distribution characteristics of Reynolds shear stress $\overline{u'v'}$ and turbulent fluctuation $u_{rms}$ in each sublayer were discussed. The law analytical expressions were established using the GDF and GEF. The mathematics and physics to determine the control parameters for each sublayer were provided based on the sublayers partition. These formulations are fundamentally governed by the identical mathematics, specifically by GDF and GEF, which guarantee the law under the two scales being consistent and the rescaling parameter being identified as $u_\tau / \overline{u_\tau} = \sqrt{1+\Delta^+}$. The formulations were validated by the existing DNS and experiment data for both incompressible and compressible TBLs and demonstrated the capability of not only representing the velocities, but also capturing the near-wall gradients. The properly scaled Type-A TBL governing equations provide the theoretical foundation for these formulations to possess the $Re_\tau$-independency property. Given the similarity of the properly-scaled governing equations, these law formulae are logically reasoned being applicable to temperature profile in thermal Type-A TBL.

**Acknowledgement**

The DNS data of zero-pressure-gradient flat-plate incompressible turbulent boundary layers were provided by Schlatter and Orlu from the KTH website at https://www.mech.kth.se/~pschlatt/DATA/README.html. The experiment data were supplied by Österlund and Nagib from KTH website at https://www.mech.kth.se/~jens/zpg/. And the compressible flow TBL data were given by Pirozzoli and Bernardini available at http://newton.dma.uniroma1.it/database/. Authors would like to sincerely acknowledge the significant support from these data and their generosity to share the data with the research community.